\documentclass{article}
\usepackage{frascatiphys}
\usepackage{graphicx}
\usepackage{multirow}
\usepackage{subfig}
\usepackage{comment}
\usepackage{xspace}
\usepackage{xcolor}
\usepackage[countmax]{subfloat}
\usepackage{amsmath}
\usepackage{mathtools}
\usepackage{graphicx}
\usepackage{slashed}
\usepackage{tikz-feynman,contour}
\usepackage{tikz,pgf}
\usepackage{braket}
\usepackage{amsfonts}
\usepackage{amsthm,scrextend,bbold}
\newcommand{\as}{\alpha_\text{s}}
\newcommand{\cf}{C_{\text{F}}}
\newcommand{\ca}{C_{\text{A}}}

\newcommand{\order}[1]{{\cal O}\left(#1\right)}

\DeclareMathOperator{\De}{d}
\newcommand{\de}{\De\!}
\newcommand{\xf}{x}

\newcommand{\xb}{x_{\text{B}}}

\newcommand{\gs}{\gamma_\text{soft}}
\newcommand{\gszero}{\gamma_\text{soft}^{(0)}}
\newcommand{\gsone}{\gamma_\text{soft}^{(1)}}
\newcommand{\cone}{C^{(1)}}
\begin{document}
\title{ 
Non commutativity between massless and soft limit in processes with heavy quarks
}
\author{
Andrea Ghira       \\
{\em Dipartimento di Fisica, Università degli Studi di Genova and INFN, Via Dodecaneso 33, 16146, Italy} %\\
%Second author's name, upper and lower case, not bold        \\
%{\em Institution Address in italics with no skipped lines}
}
\maketitle
\baselineskip=10pt
\begin{abstract}
 Processes involving heavy quarks can be computed in perturbation theory in two different ways: we can adopt a scheme in which the mass of the quark is considered only as a regulator of the collinear divergences because of the fact that the hard scale of the process is far bigger  or we can consider the quark as a massive particle. Each picture has its own advantages and drawbacks: we investigate the differences between the two approaches with particular attention to the soft logarithmic structure. We examine the origin of this difference, focusing on different processes involving the Higgs boson . Finally we perform the threshold resummation of the Higgs boson decay rate into a $b\bar{b}$ pair at NLL accuracy in the massive scheme.
\end{abstract}
\baselineskip=14pt

\section{Introduction} 
Quarks appear in the Quantum Chromo-Dynamics (QCD) lagrangian in different species, named flavours. From the point of view of strong interactions, different flavours are distinguished purely on the basis of the value of their masses. 
It is therefore natural to classify quark flavours according to their masses, compared to $\Lambda_{\rm QCD}\simeq 300\text{MeV}$. 
The masses of up, down and strange quarks, relevant for ordinary matter, are much smaller than $\Lambda_{\rm QCD}$, and can be taken to be zero for most applications in high-energy physics, on the other hand charm ($c$) and especially bottom ($b$) are heavy according to this definition. 
Heavy-flavour production cross-sections can be calculated in perturbative QCD because the mass of the $b$ and $c$ quarks sets the value of the coupling in the perturbative region and regulates collinear singularities.
In order to compute processes involving heavy flavour two main approaches are employed. In  the so-called \emph{massive scheme}, the final-state heavy quarks are considered massive particles and we can compute order by order in perturbation theory the scattering amplitude. Within this approach the kinematics is treated correctly  but calculations become cumbersome at higher and higher perturbative orders. Another drawback is that large mass logarithms which arise due to the fact that the mass of the heavy quark is far smaller than hard scale of the process spoil the convergence of the perturbative series.
Therefore another framework is employed which is the so called \emph{massless scheme}.
In the massless scheme, we treat the mass of the particle only as a regulator of the collinear divergences. Consequently we do not have control on the kinematics outside the collinear region, i.e. we consider only radiation emitted at small angle.
This approach exploits the factorization theorem: the differential cross section can be written as a convolution product of a process dependent function times a fragmentation function, which is process independent and fulfills a first order linear equation that allows us to resum the mass logarithms (DGLAP).
The initial condition of the DGLAP evolution equation is set at a scale  $\mu_0^2\simeq m^2_{c,b}\gg \Lambda^2_{\text{QCD}}$ and therefore it is in the perturbative domain and it can be determined by matching the factorisation theorem with the massive scheme.
It was determined to NLO in QCD for the $b$ quark fragmentation function in \cite{Mele:1990cw,Mele:1990yq} and to NNLO in\cite{Melnikov:2004bm,Mitov:2004du}. 
The initial condition is affected by soft logarithms, that should be resummed to all-orders too~\cite{Cacciari:2001cw,Maltoni:2022bpy}.
The main problem we want to focus on is that the structure of soft logarithms in the initial condition of the fragmentation function cannot be always recovered by the massless limit of a massive-framework calculation: this strongly depends both on the considered process and on the specific observable that is computed. We will show this particular behaviour using a simple process as an example which is the decay of a Higgs boson in a $b \bar{b}$ pair.
Secondly, we want to derive a resummed expression of the differential decay rate at NLL accuracy that fully take into account the heavy quark mass and outline also in this case the non commutativity of the massless and soft limit.
\section{Interplay between soft and massless limit in $H\to b\Bar{b}$}
\begin{figure}[t]
	\begin{center}
		\includegraphics[width=0.3\textwidth]{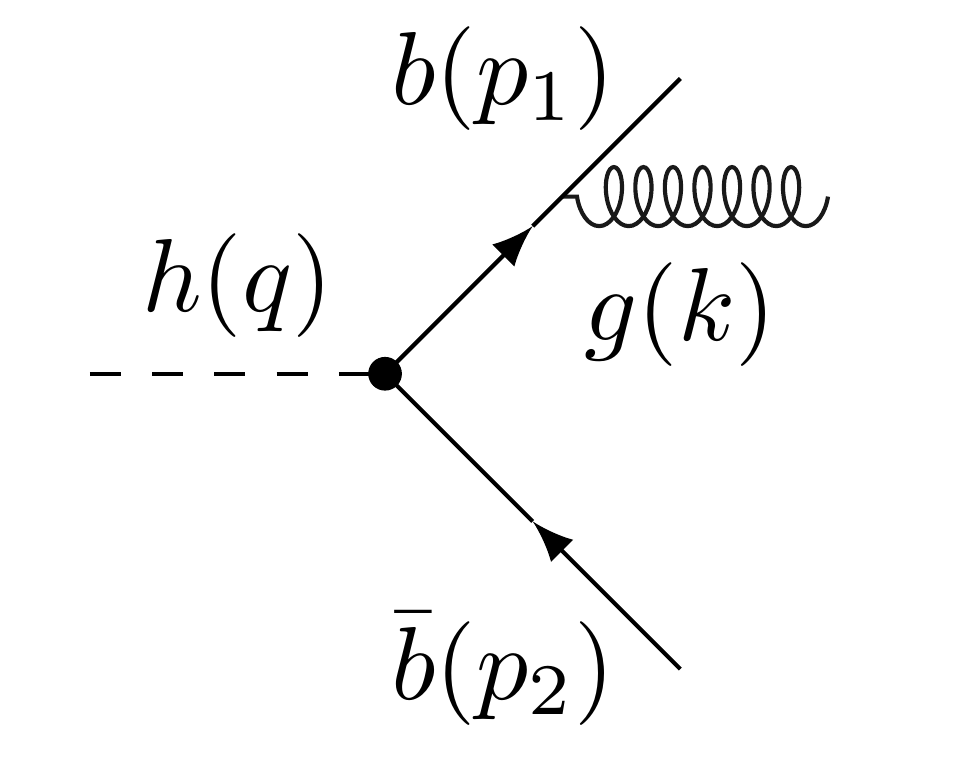}
		\includegraphics[width=0.25\textwidth]{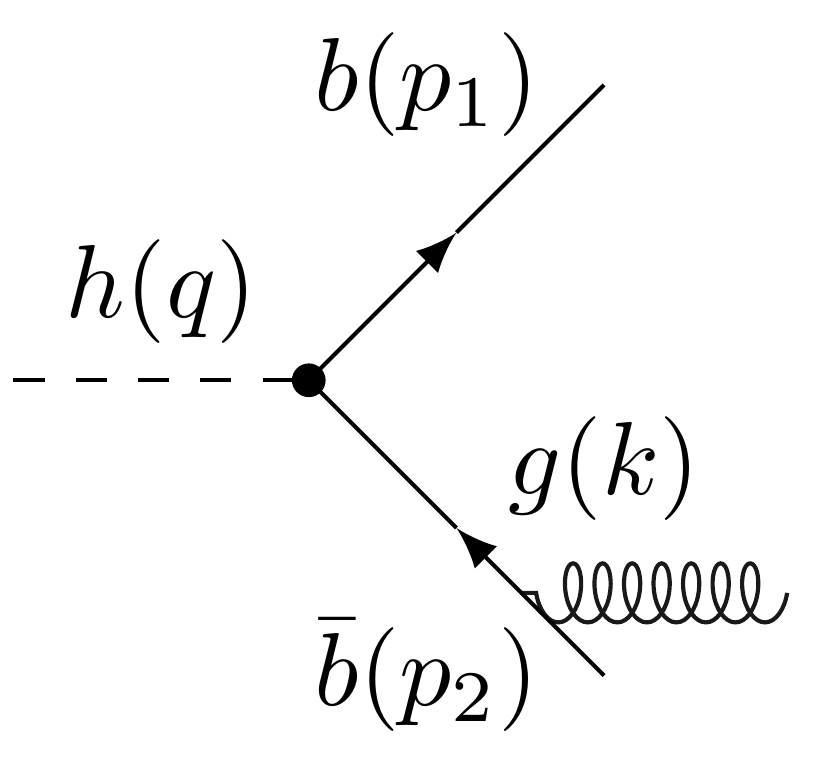}\hspace{1cm}
		\caption{\label{fig:decay}
			Real-emission contributions to the decay of the Higgs boson into a $b \bar b$ pair at $\order{\as}$.}
	\end{center}
\end{figure}
In order to explain the aforementioned non commutativity of the limits we focus on the decay of the Higgs boson at NLO keeping the mass of the quarks:
\begin{equation}
	h(q)\to b(p_1)+\bar{b}(p_2)+g(k)\qquad p_1^2=p_2^2=m^2,\; k^2=0.
\end{equation}
We compute the differential decay rate $\frac{\de \Gamma}{\de x}$, where $x=\frac{2 p_1\cdot q}{q^2}$ is the energy of the quark in the CoM reference frame, and we are interested in the small mass limit necessary for the massless scheme ($\frac{m^2}{|q^2|}\equiv \xi \to 0$) and in the soft limit ($x\to 1$). Performing the soft limit and the massless in two different orders we find:
\begin{align} 
	\label{different_limit}
	\lim_{\xi\to 0}\lim_{x\to 1}\frac{1}{\Gamma_0}\frac{\de\Gamma}{\de\xf}&= - \frac{2\as \cf}{\pi}\left[ \frac{1+\log \xi }{1-\xf} +\mathcal{O}(\xi^0)+\mathcal{O}\left((1-x)^0\right)\right],\\
	\nonumber\lim_{x\to 1}\lim_{\xi\to 0}\frac{1}{\Gamma_0}\frac{\de\Gamma}{\de\xf}
	&=-\frac{\as \cf}{\pi}\left[
	\frac{\log \xi}{1-\xf}+\frac{\log(1-\xf)}{1-\xf}+
	\frac{7}{4}\frac{1}{1-\xf}+\mathcal{O}(\xi^0)+\mathcal{O}\left((1-x)^0\right)\right],
\end{align}
where $\Gamma_0$ is the Born level decay rate:
\begin{equation}
	\Gamma_0=\frac{\sqrt{2 q^2} G_F m^2{\beta}^3 N_{\text{C}}}{8\pi}, \quad \beta=\sqrt{1-4\xi},
\end{equation}
with $G_F$ is the Fermi constant.
In order to analyze the logarithmic structure of the previous equation, we introduce the Mellin transformation:
\begin{equation}
	\mathcal{M}\{f(x)\}(N)=\int^1_0 x^{N-1}f(x)\de x
\end{equation}
We notice that in the first case of equation (\ref{different_limit}) we have a mass logarithm multiplied by a soft one ($\frac{1}{1-\xf}\leftrightarrow \log N$ in Mellin space) whereas in the second one we have an additional term which corresponds to a $\log^2 N$ after the Mellin transformation. We note also that the overall coefficient is halved in the second limit, as if the $\log(1-\xf)$ contribution in the second line of (\ref{different_limit}) is playing the role of a mass logarithm.
We would like to provide a physical interpretation to this fact:
a measurment of $x$ fixes the invariant mass $(p_2+k)^2=m^2_{g\Bar{b}}$ thus screening one of the collinear (mass) logs and preventing the anti-quark propagator to go on-shell.
In order to analyse the actual origin of the double logarithms, we have to look at the quark propagator: if we integrate it over the angle between the gluon and the quark in the $\Vec{p_2}+\Vec{k}=0$ frame we find
\begin{align}
	\int^1_{-1} \frac{1}{1-\beta_1 \cos{\theta}} \de\cos{\theta}=
	\log{\frac{\xf^2}{\xi(1-\xf)}}+\mathcal{O}\left((1-x)^0\right),\qquad \beta_1=\frac{x\sqrt{1-4\xi/x^2}}{x-2\xi},
\end{align}
where $\beta_1$ is the quark velocity in that reference frame.
In this limit, collinear logarithms appear in two distinct ways: as explicit logarithm of the quark mass $m$ or as logarithms of $1-\xf$.
This consideration brings us to formulate a more general statement about double soft logs in processes with heavy quark.
We expect this behaviour to arise if look at a differential distribution which is directly  related to the virtuality of one of the propagators, here $m^2_{g\Bar{b}}$.
Let us consider the differential distribution in $\Bar{x}=\frac{(p_1+p_2)^2}{q^2}\to 1$ as $k\to 0$. Performing an explicit calculation:

\begin{align}
	\lim_{\xi\to 0}&\lim_{\Bar{x}\to 1}\frac{1}{\Gamma_0}\frac{\de\Gamma}{\de\bar{x}}=\lim_{\Bar{x}\to 1}\lim_{\xi\to 0}\frac{1}{\Gamma_0}\frac{\de\Gamma}{\de\bar{x}}= -\frac{2 \as \cf}{\pi}\frac{1+ \log \xi}{1-\bar{x}}+\mathcal{O}(\xi^0)+\mathcal{O}\left((1-x)^0\right),
\end{align}
In this case we have only a single logarithmic enhancement and the two limits commute.

\subsection{Higgs Production and Higgs DIS}
\begin{comment}
\begin{figure}[t]
	\begin{center}
		\includegraphics[width=0.25\textwidth]{figures/Production1}\hspace{1cm}\
		\includegraphics[width=0.25\textwidth]{figures/Production2}
		\caption{\label{fig:production}
			Real-emission contributions to the production of the Higgs boson in $b \bar b$ fusion at $\order{\as}$.}
	\end{center}
\end{figure}
\end{comment}
We test our statement by studying other processes related by crossing symmetry to the Higgs boson decay, i.e Higgs boson production and Higgs DIS.
In the Higgs production $b(p_1)+\Bar{b}(p_2)\to h(q)+g(k)$, we are differential in $\tau=\frac{(p_1+p_2)^2}{q^2}$, which is not related to the virtuality of the propagators. 
In this case we find that the limits commute, as expected:
\begin{align}
	&\lim_{\tau\to 1}\lim_{\xi\to 0}\frac{1}{\sigma_0}\frac{\de \sigma}{\de \tau}= \lim_{\xi\to 0}\lim_{\tau\to 1}\frac{1}{\sigma_0}\frac{\de \sigma}{\de \tau} =- \frac{2 \as C_{\text{F}}}{\pi }  \frac{1+\log{\xi}  }{1-\tau} + \mathcal{O}(\xi^0)+\mathcal{O}\left((1-\tau)^0\right),\\
	&\nonumber \sigma_0=\frac{\sqrt{2} G_F m^2 \beta \pi N_{\text{C}}}{18 s}.
\end{align}
Finally we study the differential distribution $\frac{\de \sigma}{\de \xb}$ with $\xb=\frac{-q^2}{2p_1\cdot q}$ for the real emission corrections to the process $b(p_1)+h(q)\to b(p_2)+g(k)$.
Due to the fact that $\xb$ is related to the virtuality of one of the propagator we expect that the limit do not commute. Indeed we find:
\begin{comment}
\begin{figure}[t]
	\begin{center}
		\includegraphics[width=0.25\textwidth]{figures/DIS1}\hspace{1cm}\
		\includegraphics[width=0.25\textwidth]{figures/DIS2}
		\caption{\label{fig:scattering}
			Real-emission contributions to Higgs boson scattering off a $b$ quark at $\order{\as}$.}
	\end{center}
\end{figure}
\end{comment}
\begin{align}
	&\lim_{\xb\to 1}\lim_{\xi\to 0}\frac{1}{\bar{\sigma}_0}  \frac{\de\sigma}{\de \xb}= -\frac{\as C_{\text{F}} }{\pi}  \left[ 
	\frac{ \log\xi }{1-\xb}
	+\frac{\log(1-\xb)}{1-\xb}
	+\frac{7}{4}\frac{1}{1-\xb}
	+\mathcal{O}(\xi^0)+\mathcal{O}\left((1-\xb)^0\right)\right],\\
	&\nonumber\lim_{\xi\to 0}\lim_{\xb\to 1}\frac{1}{\bar{\sigma}_0}  \frac{\de\sigma}{\de \xb} = -\frac{2 \as C_{\text{F}} }{\pi}\frac{1+\log \xi}{1-\xb}++\mathcal{O}(\xi^0)+\mathcal{O}\left((1-\xb)^0\right),\\
	&\nonumber \bar{\sigma}_0=\frac{\pi \sqrt{2} G_F m^2 N_{\text{C}}\eta }{-3q^2},\qquad  \eta=\sqrt{1+4\xi}.
\end{align}
\section{Soft Resummation in the Massive Scheme}
In this section we want to give an explicit expression for the all-order soft resummation of the Higgs decay rate in a $b\bar b$ pair at NLL accuracy in the massive scheme. 
Since we look at the differential distribution over $x$, we are in class of process with the so called single-particle inclusive kinematics (see \cite{Laenen:1998qw}). 
The main result of \cite{Laenen:1998qw} is that the resummed expression can be factorized as a product of a soft function times a hard function times a jet function for every massles particle n the final state.  In our case the resummation formula simplifies considerably there are not massless particles.
The resummed result of \cite{Laenen:1998qw} at NLL, adapted to the process we are considering, reads\footnote{We are not so sure about the argument of the running coupling, since in \cite{Laenen:1998qw} $\as(z^2q^2)$ is used, on the other hand it seems that in \cite{Kidonakis:2009ev}  $\as(z^2 m^2)$ is used.}
\begin{align}\label{resum1}
	\widetilde{\Gamma}(N,\xi)&=\left(1+ \frac{\as}{\pi} \cone(\xi) +\order{\as^2}\right)e^{-2  \int_{1/\bar{N}}^1 \frac{d z}{z} \left[ \frac{\as(z^2 q^2)}{\pi}\gszero(\beta)+ \left(\frac{\as(z^2 q^2)}{\pi}\right)^2\gsone(\beta)+\order{\as^3}\right]} \nonumber \\&+\mathcal{O}\left(\frac{1}{N} \right),
\end{align}
with $\bar N= N e^{\gamma_E}$ and $\gs$ the massive soft anomalous dimension.
To this logarithmic accuracy we need the two loops expression of the running coupling, the coefficients $\gszero,\gsone$ and $\cone$.
The first order soft anomalous dimension can be obtained from the calculation of one gluon emission in the eikonal limit:
\begin{equation}
	\gszero(\beta)=\cf\left[\frac{1+\beta^2}{2\beta}\log{\left(\frac{1+\beta}{1-\beta}\right)}-1\right],
\end{equation}
while the second order was presented in \cite{Kidonakis:2009ev}\footnote{It is worth to mention that there is a mismatch in the literature between \cite{Kidonakis:2009ev} and \cite{vonManteuffel:2014mva}}:
\begin{align}
	\gsone&= \left\{\frac{K}{2}+\frac{\ca}{2} 
	\left[-\frac{1}{3}\log^2\frac{1-\beta}{1+\beta}+\log\frac{1-\beta}{1+\beta}-\zeta_2\right] \right.
	\nonumber \\ &% \hspace{15mm} 
	\left.
	+\frac{(1+\beta^2)}{4 \beta} \ca \left[{\rm Li}_2\left(\frac{(1-\beta)^2}
	{(1+\beta)^2}\right)+\frac{1}{3}\log^2\frac{1-\beta}{1+\beta}+\zeta_2\right]\right\} \, \gszero(\beta)
	\nonumber \\ & %\hspace{-18mm}
	{}+\cf \ca \left\{\frac{1}{2}
	+\frac{1}{2} \log\frac{1-\beta}{1+\beta}
	+\frac{1}{3} \log^2\frac{1-\beta}{1+\beta}
	-\frac{(1+\beta^2)^2}{8 \beta^2} \left[
	-{\rm Li}_3\left(\frac{(1-\beta)^2}{(1+\beta)^2}\right)+\zeta_3\right] \right.
	\nonumber \\ & \left.
	{}-\frac{(1+\beta^2)}{2 \beta} \left[\log\frac{1-\beta}{1+\beta}
	\log\frac{(1+\beta)^2}{4 \beta}-\frac{1}{6}\log^2\frac{1-\beta}{1+\beta}
	-{\rm Li}_2\left(\frac{(1-\beta)^2}{(1+\beta)^2}\right)\right]\right\},
\end{align}
with $K=\ca \left( \frac{67}{18}-\zeta_2 \right)-\frac{5n_f}{9}$.
The coefficient $\cone$ is instead process-dependent, as it receives contributions from both the end-point of the real emission and from the virtual corrections (computed in the on-shell scheme).
Writing the real emission differential decay rate as:
\begin{equation}
	\frac{\de\Gamma^{(R)}}{\de\xf}= \frac{\as\cf}{\pi} \Gamma_0^{(d)} \frac{f_\varepsilon\left(\xf,\xi,\frac{q^2}{\mu^2}\right)}{(1-\xf)^{1+2\epsilon}}, \quad \Gamma^{(d)}_0= \Gamma_0\,  \frac{ \pi^\frac{5-d}{2}}{2^{d-3}\Gamma\left(\frac{d-1}{2}\right)} \left(\frac{4 \mu^2}{q^2 \beta^2} \right)^\frac{4-d}{2},
\end{equation}
the coefficient $C^{(1)}$ can be determined using the fact that virtual corrections are proportional to $\delta(1-x)$ and the identity between distributions:
	\begin{align}
		\label{f_expanded}
		\frac{f_\varepsilon\left(\xf,\xi,\frac{q^2}{\mu^2}\right)}{(1-\xf)^{1+2\varepsilon}} &=   \delta(1-\xf)\left[-\frac{f_0(1,\xi)}{2\varepsilon}+f_0(1,\xi)\log(1-2\sqrt{\xi})-\frac{1}{2}\frac{\de}{\de \varepsilon}f_\varepsilon\left(1,\xi,\frac{q^2}{\mu^2}\right)\Big|_{\varepsilon=0} \right] \nonumber \\
		& +\frac{f_0(\xf,\xi)}{(1-\xf)_+} +\mathcal{O}(\varepsilon) \; .
	\end{align}
Summing up virtual and real contributions we obtain:
\begin{align}
	\label{C1}
	\cone(\xi) &=  \frac{\cf}{2}\Bigg\{-2 \frac{\gszero(\beta)}{\cf}\left[ -2\log{\left(1-\sqrt{1-\beta^2}\right)} + \log{\frac{m^2}{q^2}} +\log\left( \frac{1-\beta^2}{4}\right)+1 \right]  -2 \nonumber \\ &+ 2L(\beta)\left(\frac{1-\beta^2}{\beta}\right)+\frac{1+\beta^2}{\beta}\Bigg[\frac{1}{2}L(\beta)\log{\left(\frac{1-\beta^2}{4}\right)}+2L(\beta)(1-\log{\beta})+2\text{Li}_2\left(\frac{1-\beta}{1+\beta}\right) \nonumber \\ & +L(\beta)^2+L(\beta)\log{\frac{1-\beta}{2}} +\frac{2}{3}\pi^2 -\frac{1}{2}\left(\text{Li}_2\left(\frac{4\beta}{(1+\beta)^2} \right)-\text{Li}_2\left(\frac{-4\beta}{(1-\beta)^2} \right)\right)\Bigg] \Bigg\}  \; ,
\end{align}
with $L(\beta)=\log\left(\frac{1+\beta}{1-\beta}\right)$.
 We note that the non commutativity of the soft and massless limits has  consequences for the resummed expression in the massive scheme:
In the small $\xi$ limit we find:
\begin{equation*}
	\as C^{(1)}(\xi)= \as\cf\left(\frac{1}{2}\log^2{\xi}+\log{\xi}+\mathcal{O}(\xi^0)\right).
\end{equation*}
We have a double log of the mass in disagreement with DGLAP evolution equation. 
The problem is that equation (\ref{f_expanded}) does not hold if we perform the massless limit because in this limit $f_0(1,\xi)$ is not defined.
In a certain way we can say that double mass logarithms in the soft limit of the massive calculation and double soft logarithms of the massless scheme are connected.
A well defined expression in the massless limit can be obtained rewriting the differential decay rate as:
\begin{equation}
	\frac{1}{\Gamma_0}\frac{\de \Gamma}{\de\xf}= \delta(1-\xf) + \frac{\as }{\pi} \left[\cf \left(\frac{f_0(\xf,\xi)}{1-\xf}\right)_+ + A(\xi) \,\delta(1-\xf)\right],
\end{equation}
The delta coefficient has an expected behaviour for  $\xi\to 0$
\begin{equation}
	A(\xi)=\cf\frac{3}{2}\log{\xi}+\mathcal{O}(\xi^0).
\end{equation}
\section{Conclusions}
We have considered observables with different kinematics in processes involving heavy quarks, and in all processes we have computed NLO corrections taking into account the mass dependence of the square amplitude.
We have underlined that soft and massless do not always commute, in particular in the massless limit the structure of the distributions can radically change
because of the presence of double logs of $N$.
	We have traced back the origin of this particular behaviour  to the interplay between the observable we are computing and the fermionic propagators in the scattering amplitudes.
	Finally, we have focused on the massive scheme resummation of the process $H\to b\Bar{b}$ in the soft limit and we have found that within this approach double logarithms of the mass may appear, and the origin of this surprising behaviour can be lead back again to the non commutativity between the large $N$ and small mass limit.
	
An interesting phenomenological study, in the context of heavy-quark calculations, would be combine the massive scheme with the massless one where also soft logarithms are resummed.
The merging of the two becomes far from trivial because of the lack of commutativity of the limits. One would like to design an all-order matching scheme that takes into account both the different logarithmic behaviour that arises in the two cases. 
\section{Acknowledgements}
We thank Simone Marzani and Giovanni Ridolfi for the aid in the drafting of this proceeding, which is entirely based on \cite{Gaggero:2022hmv}.

\end{document}